\documentclass[12pt]{iopart}
\usepackage{graphicx}
\usepackage{breqn}
\usepackage{cite}
\usepackage{notoccite}
\usepackage{siunitx}
\usepackage[font=small,bf]{caption}

\begin{document}

\title[]{High-speed fluorescent thermal imaging of quench propagation in high temperature superconductor tapes}
\author{Roland Gyur\'{a}ki$^1$, Fr\'{e}d\'{e}ric Sirois$^2$ and Francesco Grilli$^1$}
\address{$^1$Karlsruhe Institute of Technology, Germany}
\address{$^2$Polytechnique Montr\'{e}al, Canada}
\ead{roland.gyuraki@kit.edu}

\begin{abstract}
Fluorescent Microthermographic Imaging, a method using rare-earth fluorescent coatings with temperature-dependent light emission, was used for quench investigation in high temperature superconductors (HTS). A fluorophore was embedded in a polymer matrix and used as a coating on top of an HTS tape, while being excited with UV light and recorded with a high-speed camera. Simultaneously, the tape was pulsed with high amplitude, short duration DC current, and brought to quench with the help of a localized defect. The joule heating during a quench influences the fluorescent light intensity emitted from the coating, and by recording the local variations in this intensity over time, the heating of the tape can be visualized and the developed temperatures can be calculated.
In this paper, the fluorophore Europium tris[3-(trifluoromethylhydroxymethylene)- (+)-camphorate] (EuTFC) provided sufficient temperature sensitivity and a usable temperature range from 77~K to 260 K. With the help of high-speed recordings, the normal zone development was imaged in a \SI{20} {\micro\meter} copper stabilized HTS tape held in a liquid nitrogen bath, and using a calibration curve, the temperatures reached during the quench have been calculated.
\end{abstract}
\noindent{\it Keywords\/}:{Quench, High-speed recording, Fluorescence, Thermal imaging, High temperature superconductor}
\maketitle
 
\section*{Introduction}
High temperature superconductors (HTS), are promising for electrical applications, such as motors, generators, cables and high field magnets \cite{Rey2015,Melhem2012,Kalsi2011,Luiz2011,Weijers2010}, due to their high critical current density in high applied magnetic fields and to their mechanical properties. 

However, a problem associated with HTS applications is the behaviour of the conductor when part of it reverts from the superconducting state back to its normal conducting state due to some initial disturbance \cite{Armenio2008,K2009}. This phenomena is referred to as ``quench'', and it can be caused by e.g.~excessive local temperatures, magnetic fields or non-uniform current density. After a quench, a normal conducting zone is present in the tape. This zone has a finite resistance that dissipates energy due to joule heating. Due to the thermal conductivity, this now normal zone starts to spread along the length of the tape at a given speed, referred to as the Normal Zone Propagation Velocity (NZPV). While this speed can be in the range of m/s in low temperature superconductors (LTS) \cite{Trillaud2005}, in the case of HTS, it is generally in the order of a few cm/s \cite{Pelegrin2011,Schwartz2008}.

The NZPV is a crucial aspect for the thermal stability of the tape, since a slow speed means that a small portion of the tape is bearing the full heating power, whereas a larger normal zone has an extended surface that can dissipate the heat more effectively to e.g.~a liquid nitrogen bath. Generally speaking, one would want a fast NZPV to avoid the conductor being damaged by excessive local heating. For this reason, it would be very useful to be able to monitor the temperature of the tape when designing a new device that is prone to quenching. However this has proven to be challenging to over large areas and with sufficient temporal resolution.

In this work, a new optical method is presented for the measurement of quench propagation and NZPV in HTS tapes. In addition to imaging the quench itself, the method allows measuring the spatial temperature profile on the surface of the sample. The capability of mapping the temperature distribution over a 2D surface at high speeds (millisecond time scale) constitutes the main novelty of the method. The technique is based on the temperature dependent light emission of a rare-earth fluorophore in conjunction with a high-speed camera, capable of recording thousands of images per second. Together, these allow for direct observation of dynamic events, such as the quench, in the time range of milliseconds. Using the light intensity of each pixel in the recording and adequate post-processing steps allow for the extraction of thermal data. Hence, the measurements provide both qualitative and quantitative temperature information, which can be used to compare quench behaviour of various tapes and architectures. This work shows a proof of concept of the developed method together with preliminary results of quench propagation measurements in a copper stabilized HTS tape.
 
\section{Fluorescent microthermographic imaging}
Fluorescent Microthermographic Imaging (FMI) was developed in 1982 by Kolodner et al.  \cite{Kolodner1983,Kolodner1982} as a temperature measurement method for providing high spatial and thermal resolution. It was demonstrated on MOSFETs in a packaged integrated circuit and the method showed a temperature accuracy of 0.01 $^{\circ}$C and a spatial resolution of \SI{15} {\micro\meter}. A year later, a temperature accuracy of 0.08 $^{\circ}$C and a spatial resolution of \SI{0.7} {\micro\meter} were reported. Haugen et al. \cite{Haugen2007,Haugen2008} then used FMI for the first time to image superconducting bridges carrying small amounts of currents and demonstrated that measurements are possible in superconductors down to 4.2 K.

\subsection{Absorption and emission}
In this paper, Europium tris[3-(trifluoromethyl- hydroxy-methylene)-(+)-camphorate], abbreviated EuTFC, was used for high-speed fluorescent thermal imaging \cite{Haugen2007,Haugen2008}. Fluorescence spectroscopy was done on a sample prepared as described in next section (\textit{Sample Preparation}), in order to determine the optimal illumination wavelength, as well as to confirm the emission spectrum. 

\begin{figure}[!b]
\center{\includegraphics[width=8cm]{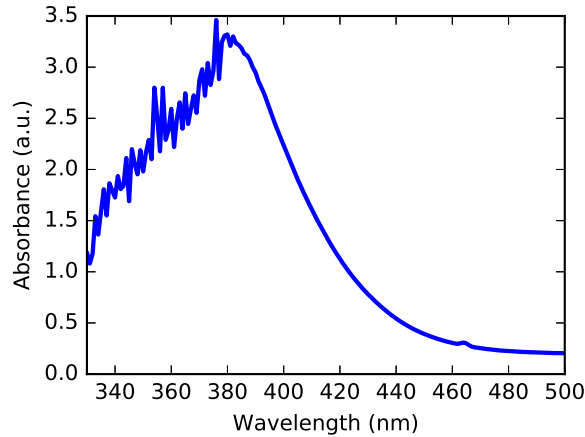}}
\caption{Absorbance spectra of EuTFC embedded in PMMA polymer film at room temperature. An absorption of over 90\% of the incoming light is possible between 340 nm and 400 nm.}
\label{fig:absorbance}
\end{figure} 

\begin{figure}[!t]
\center{\includegraphics[width=8cm]{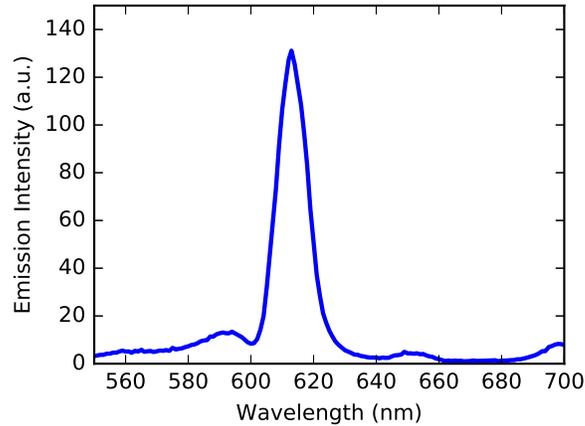}}
\caption{Emission spectra of EuTFC embedded in PMMA polymer film and excited with 365 nm light at room temperature. The principal emission peak of 612.9~nm is clearly visible.}
\label{fig:emission}
\end{figure}

The results are shown in figure~\ref{fig:absorbance} and figure~\ref{fig:emission}, where figure \ref{fig:absorbance} uses arbitrary units for showing the absorbance of the fluorophore. Absorbance is  defined by the Beer-Lambert law as
\begin{equation}
\alpha = \log (\dfrac{I_0}{I_1}) \,,
\end{equation}
where $I_0$ is the intensity of the incoming light and $I_1$ is the intensity of the light measured after passing through the sample. Hence, an absorbance of 1 means that 90\% of the light is absorbed by the sample, whereas an absorbance of 3 would mean that 99.9\% is absorbed. Over 90\% of the light between wavelengths of 360 nm and 400 nm is absorbed with the absorbance peaking at 376 nm.

Figure \ref{fig:emission} shows the characteristic emission spectra of EuTFC at room temperature and over a narrow wavelength scale, in arbitrary units as well, however here the units simply represent a digital signal without any absolute meaning. The peak emission was observed at 612.9 nm, which correlates with previous literature \cite{Haugen2008}. These results were used for selecting the optimal excitation light source and optical filters to use in the experimental setup.

\subsection{Temperature dependence}
\begin{figure}[!t]
\centering
\includegraphics[height=6cm]{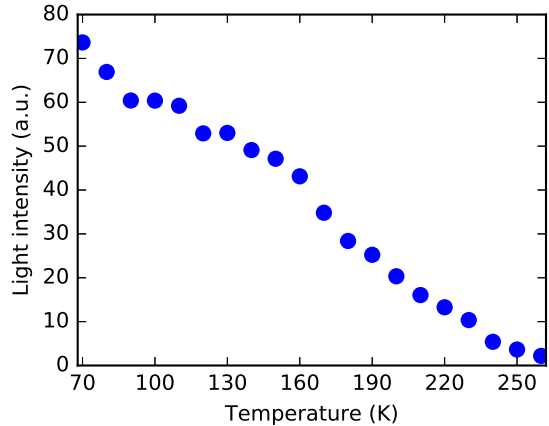} 
\caption{Light intensity (in arbitrary units) of EuTFC embedded in PMMA, baked at 175 $^{\circ}$C for 30 minutes. The range of 70~K to 260~K was investigated in steps of 10~K, which is an interesting temperature range for HTS coated conductors.}
\label{fig:EuTFC_TempvsLight}
\end{figure}

The emission peak  shown in figure \ref{fig:emission} changes its intensity depending on the temperature. It can thus be used to extract temperatures by recording the emission strength during a calibration measurement.
Figure \ref{fig:EuTFC_TempvsLight} shows temperature dependent fluorescent spectroscopy of an EuTFC sample over a temperature range from 70~K to 260~K.
One can observe a quasi-linear decrease in light intensity with increasing temperature over this whole range. Beyond 260 K the fluorescence is barely recordable, hence this constitutes the upper temperature limit of the method. Since HTS conductors are often cooled with liquid nitrogen, the temperature range from 77 K to 260~K (nearly room temperature) is perfectly convenient for the purpose of quench characterization.

In order to generate the data shown in figure~\ref{fig:emission}, the sample has been excited using a frequency doubled Ti:Saphire laser (excitation wavelength of 365 nm) and measured with an AvaSpec-ULS3648 high-resolution spectrometer, after the emission light has been filtered with a 610 nm central wavelength bandpass filter with a bandwidth of 10 nm. The temperature control was done in an Oxford closed cycle cryostat controlled by a LabVIEW code. Light intensity was measured at every 10~K by setting the required temperature. Upon reaching the target temperature, the sample was given 5 minutes to reach thermal equilibrium. At this point, the laser beam excited the sample and a measurement was taken. Following this, the laser beam's path was blocked while the sample was reaching a new measurement temperature in order to avoid photobleaching of the EuTFC layer due to prolonged excitation.

\subsection{UV excitation light}
Previous measurements have often used Hg-Xe lamps \cite{Haugen2007,Haugen2008} and Hg-arc lamps \cite{Kolodner1983,Kolodner1982,Hampel1996} in a microscopy setup as a source for excitation for fluorescence. The light of the lamp was filtered, only allowing the necessary UV component to pass. In addition, the UV lighting was directed towards the sample using an optical guide. Such optical guides, however, generally only allow a small area to be illuminated. This solution was not deemed favourable to excite an area as large as 15 cm by 1.2 cm necessary in this research. For this reason, as well as due to size and place limitations, instead of a Hg-Xe or Hg-arc lamp, UV LEDs were used in the experiment with a wavelength of 365 nm. At this excitation wavelength, more than 90\% of the light is absorbed by the fluorophore, as shown previously in figure~\ref{fig:absorbance}.

The light intensity is one of the key aspects that make such a measurement work. In static measurements, the exposure time of the camera can be arbitrarily chosen. Furthermore, additional post-processing can be applied to improve the final image quality. Haugen et al. \cite{Haugen2007} achieved 75-84 mK resolution using an exposure time of 2~s. Before that, Kolodner et al. \cite{Kolodner1983}  have used 20~s exposures and then combined the images into a total effective exposure of 320~s to improve temperature resolution to $\pm8$~mK.

Achieving the same image quality and thermal resolution in the time frame of a few tens of milliseconds is more challenging. The exposure times are limited by the transient effect to be recorded. As a consequence, the detected fluorescence is much fainter. In order to reach acceptable emission levels at frame rates between 500 and 2500 pictures per second, first a total of 4 UV LEDs (and eventually 8) were used in close proximity of the sample~\cite{LEDENGIN2016}. The total radiant flux power is therefore of approximately 4,800 to 12,000 mW, however some of this power is lost due to absorption and reflection in glass interfaces as well as in form of stray illumination (light not reaching the sample). The exact level of illumination at the surface of the HTS tape is hard to determine as it depends strongly on the distance to the sample and exact amount of DC current flowing in the LEDs. 

Note that although the optical density of liquid nitrogen is different than that of air, it was found that UV light can still propagate through lengths of tens of centimetres without considerable attenuation in this medium.

\subsection{Camera}
While fluorescent thermographic imaging has been previously used in microscopy applications (see \cite{Hampel1996,Haugen2007,Haugen2008,Kolodner1982,Kolodner1983}), observing large objects in transient state in a time scale of milliseconds posed new challenges, as already mentioned above. The main determining factors about the applicability and final quality of the recordings are:
\begin{itemize}
\item Digital resolution (in bits)
\item Recording speed (in frames per second, fps)
\item Recording resolution
\end{itemize}
Digital resolution defines the number of discrete levels available for the digitalization of the analogue light signal. Our Memrecam HX3 camera can operate with 12-bit digital resolution, meaning that 4096 different light intensity levels can be recorded at any chosen recording speed. This directly defines the maximum theoretical temperature resolution of the measurement. If the maximum light intensity of the CCD chip, i.e. the value 4095, is achieved at the temperature of 77~K, and if the sensor records a value of 0 at the maximum temperature (260~K), then the theoretical maximum temperature resolution is,
\begin{center}
$(260-77) / {(4095-0)}=0.0447$ K.
\end{center}
This is, however, unrealistic, mainly due to noise in the CCD sensor, which reduces the effective digital resolution.

Sufficiently high recording speed and resolution are required for capturing the quench itself and having a sharp enough image. From experience, a recording speed of 2500 fps (one image every 0.4 ms) allows clearly seeing the quench propagation in HTS tapes. Anything beyond that is not strictly necessary, however it can be useful for obtaining smoother recordings. The HX3 camera can maintain an HD resolution of $1280 \times 720$~pixels at 10,000 fps; hence the speed and resolution are not limiting the accuracy of the measurement. Nevertheless the resolution of the area of interest (only the tape) is not necessarily available in high-definition due to the used lens and optics of the setup. Being at the minimum working distance of the current camera lens, the recorded images are approximately $200 \times 500$~pixels, or roughly 1~Megapixel.

Additionally a 610~nm central wavelength bandpass filter with a 10~nm bandwidth was used to filter out any light not resulting from the temperature-dependant fluorescence. Since the central wavelength of such filters shifts with the light incidence angle, mounting the filter in front of the camera lens (as in traditional photography) was unfavourable. Furthermore, these filters are made in small diameters and are mainly used for experiments in traditional optics. Here, a custom 3D printed frame was made for the bandpass filter, which could hold it between the lens and the camera's standard C-mount. This allowed all the light to first pass through the camera's lens, and then through the filter. This is illustrated in figure~\ref{fig:schematic}, described in details in the next section.

\section{Experimental details}
\subsection{Sample preparation}
A 12 mm wide HTS tape sample from SuperPower was cut into a final length of 15 cm. The fluorophore was mixed with Poly(methyl methacrylate) (PMMA) and dissolved in acetone in a ratio of 1.3~wt\%, 1.7~wt\% and 97~wt\% of fluorophore, PMMA and acetone, respectively \cite{Barton2013,Haugen2008}. A high doping of fluorophore in the PMMA matrix is required for sufficient light emission during excitation, especially when recording at high speeds. 

\begin{figure}[!t]
\centering
\includegraphics[width=8cm]{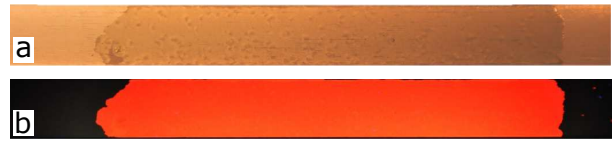}
\caption{A: shows the coating over the central section of a 15 cm long, non-stabilized HTS under normal lighting conditions. B: shows the same tape when excited with 365~nm UV light.}
\label{fig:tape}
\end{figure}

The solution, which is mostly acetone, is brought onto the surface of a sample tape using simple droplet deposition. As the acetone covers the tape and forms a thin layer, it evaporates rapidly and leaves behind a layer of fluorophore and PMMA mixture. Finally, in order to mechanically stabilize the layer and the photoluminescence itself, the sample is heat treated at 175 $^{\circ}$C for 30 minutes \cite{Haugen2008}. The heating profile was chosen arbitrarily with a rising rate of 175 $^{\circ}$C/hour, a hold time of 30 minutes at the set temperature, and natural cool down thereafter. Figure~\ref{fig:tape}a shows a 15 cm long non-stabilized HTS tape after the heat treatment, where the coating is visible over the middle section of the tape. Figure~\ref{fig:tape}b shows the same HTS tape when excited with UV light; the light emission is uniform around the characteristic 612 nm wavelength since the sample was held at constant temperature for this measurement.
  
\begin{figure}[!t]
\centering
\includegraphics[width=8cm]{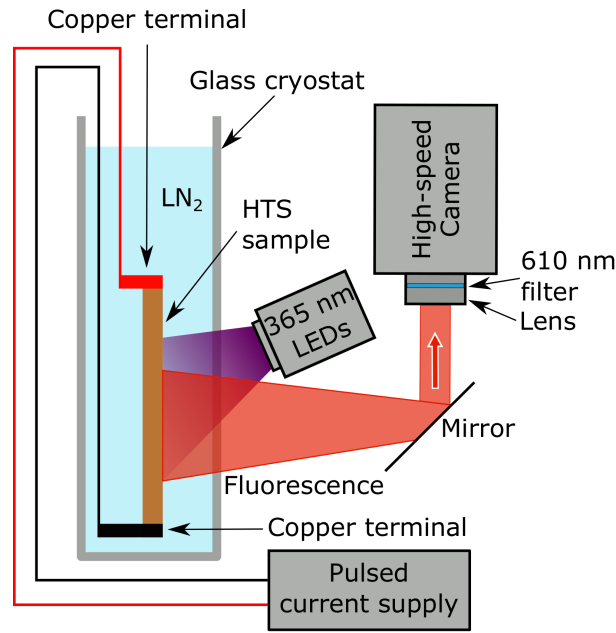}
\caption{Schematic image of the measurement setup. A sample HTS tape is placed in an optical liquid nitrogen cryostat and excited with UV light. A recording is made at high speed and at the same time a high amplitude current pulse is sent through the tape. The change in light intensity emitted by the tape indicates the surface temperature.}
\label{fig:schematic}
\end{figure} 

It has been previously shown that a thin layer of insulator between an HTS tape and a cooling bath can influence the heat transfer properties \cite{Hellmann2014,Rubeli2015}. Therefore, the presence of a fluorescent coating may have to be kept in mind when interpreting measurement results about the dynamics of the quench.

\subsection{Experimental setup}

A schematic of the experiment is shown in figure~\ref{fig:schematic}. A tape is held between two copper terminals and placed into an optical cryostat, filled with liquid nitrogen. The UV light array is placed outside of the cryostat and the light excites the tape at an oblique angle. A short, generally 50 ms to 150 ms current pulse, is sent through the tape and at the same time, the high-speed camera is triggered.
In the current setup, the sample is in a vertical position in the cryostat, which itself is placed in a larger metal enclosure covered with insulating blankets, mainly to block out any ambient light that would affect the measurement. Due to the arrangement of the metal enclosure, the camera is looking at the sample through a mirror.

\subsection{Post-processing}
Since the amount of light varies based on the temperature of the sample, recording these changes allows calculating the temperatures, based on the previously determined calibration curve. If the temperature calibration is done on the exact same setup as the measurement, with the exact same conditions, then the measurement results could be simply interpolated using the calibration curve to extract the exact temperatures. However, this is generally not the case, as it is rather difficult to accurately control the temperature of the sample for calibration in the same setup where the measurements will be performed. Furthermore, if not recorded by the same camera, the arbitrary units of measurements cannot be compared directly.

A solution to this problem was introduced by L. D. Barton \cite{Barton2013} by working with relative light intensity changes instead of using an absolute scale. Let $S(x,y)$ represent the light intensity at a point in a 2D image. This value depends on the: excitation intensity, $I(x,y)$, the optical collection efficiency, $\eta(x,y)$, the sample reflectivity, $r(x,y)$ and the quantum efficiency $Q(T(x,y))$. The relation can be described as
\begin{equation}
S(x,y)=I(x,y)\cdot \eta(x,y)\cdot r(x,y)\cdot Q(T(x,y)) \,.
\end{equation}
In order to remove all optical artefacts, one can create a relative scale instead of the absolute measurement and divide the hot (biased) image, $S_{\rm H} (x,y)$, by the cold (reference) image, $S_{\rm C} (x,y)$, to get the ratio $S_{\rm R} (x,y)$, i.e.

\begin{dmath}
S_{R}(x,y)= \frac{S_{\rm H}(x,y) }{S_{\rm C} (x,y)}
= \frac{I(x,y)\cdot \eta(x,y)\cdot r(x,y)\cdot Q(T_{\rm H}(x,y))}{I(x,y)\cdot \eta(x,y)\cdot r(x,y)\cdot Q(T_{\rm C}(x,y))}=\frac{Q(T_{\rm H} (x,y))}{Q(T_{\rm C} (x,y))} \,.
\end{dmath}
Since only the quantum efficiency of the fluorophore depends on the temperature, all other optical artefacts cancel out.  Hence, the local variations in e.g.~optical collection efficiency or excitation light intensity are no longer a concern. This allows the calibration to be done on a different measurement device as the actual measurement.

\begin{figure}[!t]
\centering
\includegraphics[height=6cm]{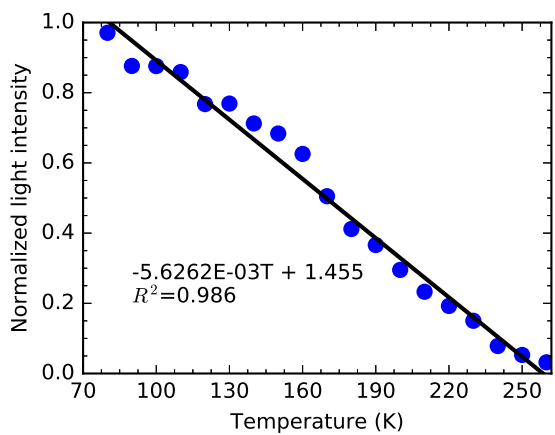}
\caption{Light intensity of EuTFC embedded in PMMA and baked at 175$^{\circ}$C and normalized around 77 K. The $R^2$ value shows a nice fit to a linear approximation.}
\label{fig:EuTFC_TempvsLight_norm}
\end{figure}

The complete calibration process works as follows: firstly, the calibration curve is normalized around 77~K as shown in figure~\ref{fig:EuTFC_TempvsLight_norm}, resulting in a relative scale. Then, the quench recording is also normalized on a pixel-by-pixel basis by dividing all images of the video with the average of the first few images, where the tape is known to be at constant temperature (here 77~K since it is immersed in a liquid nitrogen bath, but the the tape could be held at any other reference temperature). Since the light intensity in the case of this fluorophore varies almost linearly with the temperature, thermal data can be extracted from the images using linear interpolation on the normalized calibration curve.

\section{Results}

\begin{figure}
\centering
\includegraphics[height=5cm]{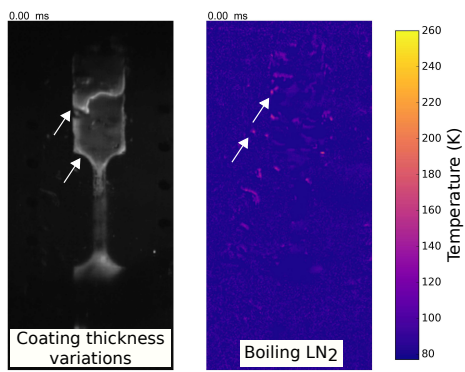} 
\caption{Reference images with a \SI{20} {\micro\meter} copper stabilized sample with a 3 cm long section at the centre narrowed down to 3 mm width for quench initiation. Left: Image of the high-speed camera where the bright layer is the fluorescence at 77 K. Right: Post-processed image using the previously shown normalization method and temperature calibration curve.}
\label{fig:ref_cam_postproc}
\end{figure}

Figure~\ref{fig:ref_cam_postproc} shows an actual image recorded by the high-speed camera next to a post-processed image following the procedure introduced above. Non-uniformities in light intensity, caused most likely by the varying thickness of the coating over the surface, are clearly visible in the camera's image. However they completely disappear in the post-processed images thanks to the normalization. On the other hand, rising  $\rm N_2$ gas bubbles due to the boiling cryogenic bath, pose a problem. Since these change position in successive images, they cannot be filtered out from the recordings and are clearly visible.

\begin{figure*}[t]
	\centering
	\includegraphics[height=5cm]{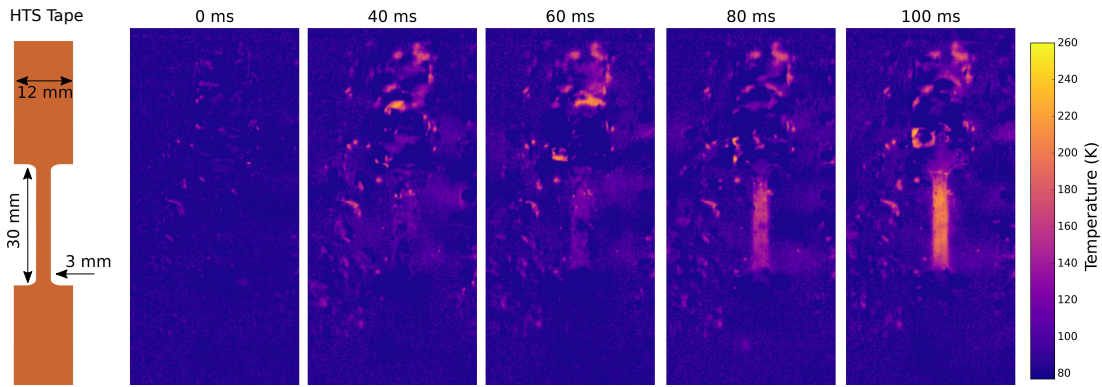}
	\caption{Schematic of the structured \SI{20} {\micro\meter} copper stabilized SuperPower tape and the time evolution of a quench following a 100 ms long rectangular current pulse with an amplitude of 250 A. The recording was done at 2,500 images per second with the sample placed in a liquid nitrogen cryostat in a vertical position.}
	\label{fig:quench_development}
\end{figure*}

Figure \ref{fig:quench_development} shows the time evolution of a quench in a SuperPower,  \SI{20} {\micro\meter} copper-stabilized HTS tape coated with EuTFC. In the current experiment, a tape is shown with a special structure, solely for controlling the quench initiation. The tape's central section has been narrowed down to 3~mm in width over a total length of 3~cm using a Nd:YAG laser. The tape was quenched using a current pulse amplitude of 250~A with a duration of 100~ms. The thermal imaging shows that the narrowed section heats up gradually, almost completely uniformly. The first heating effects are visible after 40~ms, with parts of the tape reaching close to 90~K, i.e. close to the critical temperature of YBCO conductors. At 60~ms, the narrow section is almost uniformly above the critical temperature and heats further during the current pulse. After the 100~ms mark, the tape reaches temperatures of 260~K and more, beyond which the fluorophore used here does not emit anymore light (see figure~\ref{fig:EuTFC_TempvsLight_norm}), so the colour remains apparently “frozen”, even if the temperature continues to rise in this case. The individual images in figure  \ref{fig:quench_development} are extracted from the high-speed fluorescent recording attached to this publication.

\section{Conclusion}
In this paper, we introduced a high-speed fluorescent thermal imaging technique aimed at measuring temperatures in HTS tapes during a quench.
Using a high-speed camera and EuTFC as a fluorophore, quench temperatures could be observed at high speeds in HTS coated conductors immersed in liquid nitrogen. Recording speeds of 2,500 frames per second were found sufficient, as well as UV LED arrays with a total light emission of approximately 12,800 mW placed up to 30 cm away from the sample. With EuTFC and sufficient calibration, the temperature range from 77 K to 260 K can be imaged with reasonable accuracy. It was found that liquid nitrogen does not block considerable amounts of UV light, hence the method works well even in this medium, despite the slight distortion induced by bubbles.

In a copper coated HTS tape, a gradual overall heating was observed over a 3~cm by 3~mm section. This indicates a strong current sharing between the superconducting layer and the copper coating. During the current pulse, the tape has reached the limit of temperature sensitivity, 260 K, in about 60 ms, and heated even further.

\section{Acknowledgements}
The authors would like to thank Nicolo Baroni and Marius Jakoby from the Institute of Microstructure Technology, KIT, for their help with the fluorescent spectroscopy and calibration measurements, as well as to Rainer Nast from the Institute of Technical Physics, KIT, for preparing the HTS sample.

\pagebreak
\section{References}
\bibliographystyle{ieeetr} 
\bibliography{references}

\end{document}